\documentstyle[aps,eqsecnum,preprint]{revtex}
\begin{document}
\preprint{MSUCL-997}

\title{Nuclear Flow in Consistent Boltzmann Algorithm Models}

\author{G. Kortemeyer, F. Daffin, and W. Bauer}
\address{
NSCL/Cyclotron Laboratory, Michigan State University, East Lansing, MI
48824-1321, U.S.A.}
\date{\today}
\maketitle

\begin{abstract}
We investigate the stochastic Direct Simulation Monte Carlo
method (DSMC) for numerically solving the collision-term in heavy-ion
transport theories of the Boltzmann-Uehling-Uhlenbeck (BUU) type.
The first major modification we consider is changes in the collision
rates due to excluded volume and shadowing/screening effects (Enskog theory).
The second effect studied by us is the inclusion of an additional advection
term.  These modifications ensure a non-vanishing second virial and change the
equation of state for the scattering process from that of an ideal gas to that
of a hard-sphere gas.
We analyse the effect of these modifications on the calculated value of
directed nuclear collective flow
in heavy ion collisions, and find that the flow slightly increases.

\tiny

EMail: kortemeyer@theo03.nscl.msu.edu


Electronic Manuscripts:
A LaTeX-version of the text and PostScript-files of the figures are available
and can be received via email.
\end{abstract}

\narrowtext

\section{Introduction}
One of the still most challenging questions in nuclear physics is that of the
equation of state (EOS) of nuclear matter \cite{EOS}.
To investigate the properties of
nuclear matter at high densities and temperatures, heavy ion collisions are of
great importance, and various experiments are conducted for that purpose. The
simulation of such collisions however is equivalent to solving a
quantum-mechanical many-body problem, which to date is not fully possible.
Different approaches have been made to nevertheless approximate the solution.
One method is that of Molecular Dynamics (see for example
\cite{moldyn}),
in this method both the long-range attractive (soft) and the short-range
repulsive (hard) part of the particle interaction is parametrized by
potentials, the trajectories are continously updated in response to the local
potential.

Another semi-classical particle-based method is the BUU approach. Here the
soft part of the
interaction is represented by mean fields, while the hard part is given by an
explicit collision term. The collision term itself can again be represented in
different ways, especially the criteria for a collision to happen are model
dependent. In many codes, this decision is based on geometrical considerations,
for example, a collision is generated at the point of closest approach between
two particles (see for example \cite{buu,bauer}).

One particular implementation of the collision term in BUU codes is the Direct
Simulation Monte Carlo approach (DSMC), see for example Lang. et al.
\cite{lang}, and Danielewicz \cite{bertsch}. In this approach, collisions
are not generated through geometrical and particle-trajectory based criteria,
but stochastically in a way that the correct collision rate is reproduced.
In a collision only momenta and energy of the particles are
changed, while the particles themselves stay in place until the next advection
step -- the particles are assumed to be pointlike.

It was suggested that in order to reproduce a Hard-Sphere Boltzmann Equation,
the DSMC approach should be extended by
an additional advection that should take place after any collision
\cite{alexander}, and by a modification the collision probability itself
\cite{alexander,resibois}.
This advection is supposed to push the collision partners away from each other
according to their hard-sphere radius, and the collision probability is
adjusted
to take into account the excluded volume of the hard spheres and screening
effects. The modified DSMC method is called Consistent Boltzmann Algorithm
(CBA).

In this paper we study the effect of those modifications on the nuclear
flow. In section \ref{sec2} we describe
the theoretical background of our calculations, and section \ref{sec3} presents
our results and conclusions.
\section{Theoretical Background}\label{sec2}
In the DSMC approach, the positions and momenta of the
particles are evolved in a
two-step process, namely advection and collisions, corresponding to one
timestep of the simulation. During the advection step
the particles are propagated according to their momenta. During the collision
step first the particles are sorted into
spatial cells of volume $V$. Then out of the $n$ particles within a given box,
at random, $m$ combinations are chosen and scattered with the probability
\begin{equation}
W=\frac{\sigma(\sqrt s)v_{\mbox{\scriptsize rel}}\Delta t}{NV}
\frac{n(n-1)/2}m\ ,\label{ws}
\end{equation}
where $\sigma(\sqrt s)$ is the energy-dependent elementary hadron-hadron
cross section, $N$ is the number of testparticles representing one nucleon
in a full-ensemble testparticle algorithm \cite{welke},
$\Delta t$ is the timestep length, and $v_{\mbox{\scriptsize rel}}$ is the
relative velocity of the particle pair \cite{lang}. In the limit $V\to0$,
$\Delta t\to0$, $N\to\infty$, the solutions of this method have been shown to
converge to the exact solution of the Boltzmann equation \cite{badovsky}.

This approach does not take into account the finite size of the nucleons; the
testparticles are point-like, and if it was not for the contribution of the
mean
field, they would be following an ideal gas equation of state.
It has therefore been suggested by Alexander et al. \cite{alexander} to
include an extra displacement {\boldmath $d$\unboldmath} of the collisions
partners,
\begin{equation}
\mbox{\boldmath $d$\unboldmath}=\frac12
\frac{
\mbox{\boldmath $v$\unboldmath}_r'-
\mbox{\boldmath $v$\unboldmath}_r}{|
\mbox{\boldmath $v$\unboldmath}_r'-
\mbox{\boldmath $v$\unboldmath}_r|}\sqrt{\frac{\sigma(\sqrt s)}\pi}\ ,
\end{equation}
$\mbox{\boldmath $v$\unboldmath}_r=\mbox{\boldmath $v$\unboldmath}_1-
\mbox{\boldmath $v$\unboldmath}_2$ being the velocity difference before, and
$\mbox{\boldmath $v$\unboldmath}_r'=\mbox{\boldmath $v$\unboldmath}_1'-
\mbox{\boldmath $v$\unboldmath}_2'$ being the velocity difference after the
collision. Particle 1 is displaced by {\boldmath $d$\unboldmath} and particle 2
by {$-$\boldmath $d$\unboldmath}. This additional advection pushes the
nucleons apart according to their hard-sphere radius. It is not
obvious right away how this displacement scales with $N$. However, as in the
mean free path $1/((\sigma/N)(N\varrho))$ of a testparticle, $\varrho$ being
the nuclear density, there is no $N$
dependence, the average number of collisions that a certain testparticle is
involved in is independent of $N$. Therefore, in order to achieve the
same total displacement during the
course of the simulation, the individual displacement per collision should not
depend on $N$ either. The testparticles are therefore pushed apart according to
the nucleonic radius, and not according to the effective testparticle radius.

The finite radius of the particles also makes it impossible for one
particle to be within the ``spheres of influence'' of the others, and thereby
from the available volume $V$ a fraction
\begin{equation}
\frac nN\cdot \frac43\pi a^3=
V\varrho\cdot\frac43\pi a^3
\end{equation}
is occupied, where $a$ is the average radius of the ``sphere of influence''
of one nucleon, and $n/N$ is the number of nucleons in the respective box.
We obtain $a$ by randomly picking $n$
particle combinations for a respective box and calculating their cross sections
$\sigma_i(\sqrt{s_i})$.
The Pauli principle is approximately taken into account by only choosing
particle combinations with partners that are not from the same nucleus, unless
at least one of them had scattered before, and for the remaining combinations
taking into account the reduced phase space volume. With $p_B$ being the
momentum of the beam per nucleon, and $p_F$ being the Fermi momentum, we obtain
for $a$ in this approximation
\begin{equation}
a=\frac12\sqrt{\frac1\pi\left(\frac1n\sum_{i=1}^n\sigma_i(\sqrt{s_i})\right)
\cdot\left(1-2\frac{p_F^3}{(p_F+p_B)^3}\right)}\ .
\end{equation}
For small $E_{\mbox{\scriptsize lab}}$, this effective radius is about 0.84 fm,
in the range
between 100-400 MeV it is about 0.47 fm. This is slightly larger than what has
recently been suggested in Ref.~\cite{pratt}, there, the effective radius
derived from delays in elementary processes is about
0.6-0.8 fm and 0.15-0.3 fm, respectively.

Due to the reduced volume
\begin{equation}
\widetilde V=\left(1-\varrho\cdot\frac43\pi a^3\right)V
\end{equation}
alone, a modified higher scattering probability
\begin{equation}
\widetilde W=\frac V{\widetilde V}W
\end{equation}
has to be used. However, the scattering probability is lowered again by
another effect: the particles are screening each other. A particle might not be
available for scattering with another particle because there might be a third
particle in between. It can be shown \cite{resibois}, that including this
effect leads to a reduction of the scattering probability by a factor of
\begin{equation}
\left(1-\varrho\cdot\frac{11}{12}\pi a^3\right)\ .
\end{equation}
Again, the product $\varrho\cdot a^3$ is independent of the number of
testparticles per nucleon $N$. Including this factor, the modified scattering
probability is
\begin{equation}
W'=Y^EW\ ,
\end{equation}
where
\begin{equation}
Y^E=\frac{1-\varrho\cdot\frac{11}{12}\pi a^3}{1-\varrho\cdot\frac43\pi a^3}=
\frac{1-11 b^E\varrho/8}{1-2b^E\varrho}\ ,\quad
b^E=\frac23\pi a^3\ ,\label{ye}
\end{equation}
$b^E$ being the second virial coefficient as yielded
by the Enskog Theory of the dense hard-spheres fluid \cite{resibois}; see
figure \ref{fig1}.
The implementation of the modifications makes the second virial of the hard
part of
the interaction non-vanishing; $b^E$ is positive and therefore leads to an
increase in preasure. This is partly compensated by the negative virial $b^S$
that is due to the soft (mean field) part of the interaction; see for example
\cite{pratt}.
The equation of state deviates from that of an ideal gas by both contributions,
i.e.,
\begin{equation}
P=\varrho kT\left(1+\varrho(b^E+b^S)+\ldots\right)\ .
\end{equation}
One should note at this point that the Enskog Theory is non-relativistic;
both the advection vector {\boldmath $d$\unboldmath}
and the excluded volume, therefore also $Y^E$, are calculated in a
frame-dependent way.
\section{Results and Conclusion}\label{sec3}
Our numerical calculation is based on the MSU BUU-code by Bauer et al.
\cite{bauer} which was modified from a geometrical formulation
of the collision term to a stochastic formulation according to
Ref.~\cite{lang}. Only $NN$
collisions were taken into account, which for the energies considered turned
out to be a justified approximation. A full-ensemble
and a parallel-ensemble implementation proved to have similar results;
results for different systems were compared with both
Refs.~\cite{bauer} and Ref.~\cite{danielewicz}; they were
found in satisfactory agreement. An interesting side-result at this stage
however was that changing the algorithm from geometrical to stochastic
scattering increased the frame-of-reference dependence of the result: when
running the simulation within
the lab-frame we found an asymmetry in the flow which we attribute to the
fact that in this frame the projectile is Lorentz-contracted and the
target is not.
Therefore within a spatial box inside of the overlap-zone of the two nuclei
there are many more testparticles originating from the projectile than from the
target, resulting in an asymmetry of the respective scattering rates. We are
currently trying to overcome this problem by $\gamma$-dependent modifications
of the scattering probabilities, however, so far with only little success.
The flow-asymmetry vanishes when the calculation is performed in the
c.m.-frame, which is what we did in this work. The geometry-based code did not
appear to be sensitive to this asymmetry, however, Lorentz invariance certainly
still is an issue \cite{kortemeyer}. Finally the full-ensemble version of the
stochastic code was modified according to Ref.~\cite{alexander}, and the
scaling of the collision rate and the flow with the number of testparticles
$N$ was checked.

We simulated an (Au,Au)-collision at projectile energies
of 250 and 400 MeV, and $b=3$ fm over a total time of 70 fm/$c$. The timestep
length was 0.1 fm/$c$, the volume $V$ was approximately 1.8 fm$^3$, the
number of testparticles per nucleon $N$ was 250, $m$ in equation (\ref{ws}) was
chosen to be $n^2$, and we used a soft momentum-dependent equation of state.
Figure \ref{fig2} shows the evolution of the configuration in the reaction
plane, plotted is the nuclear density $\varrho(x,y=0,z)$.
The result for the unmodified algorithm is shown on the left, the result for
the modified algorithm on the right, respectively. As expected, the nuclei
disintegrate slightly more violently due to the additional advection, resulting
in a lower nuclear density. A calculation with the
additional advection alone revealed that therefore also the collision rate
decreases. However, this is partly compensated by the modified
scattering probability Eq.~(\ref{ye}). On the other hand, a calculation with
the modified scattering probability alone shows a strongly enhanced collision
rate, as expected. The upper panel of figure \ref{fig3} illustrates this for
the 250 MeV collision, the lower panel shows the collision rates for the 400
MeV collision.

Figure \ref{fig4} shows the average final transverse momentum
versus the reduced rapidity as an indicator for nuclear flow.
{}From the slope at zero transverse momentum one concludes that in this
specific
simulation the flow increases by approximately 16\% for the 250 MeV
collision,
and 5\% for the 400 MeV collision, with the introduction of the new algorithm.
Introducing only the additional advection, as already pointed out, the
collision rate slightly decreases, however, the nuclear flow increases by about
8\% for the 250 MeV collision. With the introduction of the modified scattering
probability alone, the flow increases by about 13\%. The fact that the
contributions from both modifications do not ``add up'' indicates that a
perturbative approach to their representation would not be justified.

An impact parameter averaged analysis for 250 MeV collisions with $b\le5$ fm
resulted in approximately 132 MeV/($c\cdot$Unit of Reduced Rapidity) nuclear
flow for the  unmodified, and 148 MeV/($c\cdot$Unit of Reduced Rapidity) for
the modified algorithm (12\% increase). The Plastic Ball data indicate
approximately 130 MeV/($c\cdot$Unit of Red. Rap.) \cite{plastic} nuclear flow,
the EOS data only 119 MeV/($c\cdot$Unit of Red. Rap.) \cite{EOSnew}.
For 400 MeV collisions the same calculations resulted in
in approximately 166 MeV/($c\cdot$Unit of Red. Rap.) nuclear
flow for the  unmodified, and 185 MeV/($c\cdot$Unit of Red. Rap.) for
the modified algorithm (11\% increase; Plastic Ball: $\approx$ 169 MeV/
($c\cdot$Unit of Red. Rap.) \cite{plastic}; EOS:
$\approx$ 151 MeV/($c\cdot$Unit of Red. Rap.) \cite{EOSnew}).

Overall we found the effect of the additional advection and the modified
scattering probability to be significant, but not crucial. Their implementation
moves the outcome of the simulations away from the experimental results. This
indicates the need for an in-medium reduction of the $NN$ cross section. This
type of reduction was first found to be needed in studies of the disappearance
of flow \cite{westfall} and later also in theoretical studies based on
thermodynamic $T$-matrix theory at finite temperature \cite{alm}. These results
were obtained by algorithm with closest approach techniques. If one wishes to
address the question of the nuclear equation of state with a DSMC algorithm,
however, the corrections discussed in the present paper should be taken into
account.
\acknowledgments
We acknowledge useful discussions with P. Danie\-lewicz.
Research supported by an NSF presidential faculty fellow award and by
NSF grants 9017077 and 9403666, and by the Studienstiftung des Deutschen
Volkes (GK).

\begin{figure}
\caption[]{The scattering enhancement factor $Y^E$ as a function of the
average cross section. Shown is $Y^E$ for $\varrho=1,2,3\varrho_0$.
\label{fig1}}
\end{figure}

\begin{figure}
\caption[]{The nucleon configuration in the reaction plane at
different times during the collision, shown is $\varrho(x,y=0,z)$.
The two columns of panels on the left
refer to the 250 MeV collisions, the two columns on the right to the 400 MeV
collisions.
Within those columns the respective panels on the left result from an
unmodified simulation with point-like nucleons, the panels on the right from a
hard-sphere simulation according to Ref.~\protect\cite{alexander}. The
difference is hardly visible, even though from the two latest panels one
gets the impression that the nuclei disintegrate slightly more violently
with the new algorithm.
\label{fig2}}
\end{figure}

\begin{figure}
\caption[]{Collision rate for a 250 and 400 MeV (Au,Au) collision with
$b=3$ fm versus time. The solid curve refers to the unmodified simulation, the
dashed curve to the modified one. For the 250 MeV collision two dotted lines
were added, the upper line refers to a calculation where only the scattering
enhancement was taken into account, the lower line to a calculation that only
incorporated the additional advection.
\label{fig3}}
\end{figure}

\begin{figure}
\caption[]{Average final transverse momentum versus reduced rapidity of the
protons in a 250 MeV (upper panel) and 400 MeV (lower panel) (Au,Au) collision
with $b=3$ fm. The reduced rapidity is the rapidity $Y_{\mbox{\scriptsize cm}}$
devided by the rapidity of the beam, which for the 250 MeV collision is
approximately 0.36, and for the 400 MeV collision approximately 0.45.
{}From the slope of a linear fit around the origin, one can determine the
nuclear flow. The circles and the solid fit refer to the unmodified,
the stars and the dashed fit to the modified algorithm. For the 250 MeV
collision, the flow is
$\approx$ 147 MeV/($c\cdot$Unit of Red. Rap.) for the unmodified,
and $\approx$ 170 MeV/($c\cdot$Unit of Red. Rap.) for the modified
algorithm. A calculation that only took into account the scattering
enhancement resulted in $\approx$ 166 MeV/($c\cdot$Unit of Red. Rap.), a
calculation that only incorporated the additional advection in
$\approx$ 159 MeV/($c\cdot$Unit of Red. Rap.).
For the 400 MeV collision, the flow for the unmodified algorithm is
$\approx$ 185 MeV/($c\cdot$Unit of Red. Rap.); for the modified
algorithm it is $\approx$ 195 MeV/($c\cdot$Unit of Red. Rap.).
\label{fig4}}
\end{figure}
\end{document}